\documentclass[twocolumn]{aastex631}

\usepackage[utf8]{inputenc}
\usepackage{textcomp}
\usepackage{upgreek}
\makeatletter
\DeclareRobustCommand{\SiXI}{%
  \mbox{Si \check@mathfonts\fontsize\sf@size\z@\selectfont XI}%
}
\makeatother
\makeatletter
\DeclareRobustCommand{\SiX}{%
  \mbox{Si \check@mathfonts\fontsize\sf@size\z@\selectfont X}%
}
\makeatletter
\DeclareRobustCommand{\SiIX}{%
  \mbox{O \check@mathfonts\fontsize\sf@size\z@\selectfont III}%
}
\makeatother
\makeatletter
\DeclareRobustCommand{\SiVI}{%
  \mbox{N \check@mathfonts\fontsize\sf@size\z@\selectfont II}%
}
\makeatletter
\DeclareRobustCommand{\FeXIII}{%
  \mbox{Fe \check@mathfonts\fontsize\sf@size\z@\selectfont XIII}%
}
\makeatother
\makeatletter
\DeclareRobustCommand{\FeVI}{%
  \mbox{Fe \check@mathfonts\fontsize\sf@size\z@\selectfont VI}%
}
\makeatletter
\DeclareRobustCommand{\AlIX}{%
  \mbox{Al \check@mathfonts\fontsize\sf@size\z@\selectfont IX}%
}
\makeatother
\makeatletter
\DeclareRobustCommand{\AlVI}{%
  \mbox{Al \check@mathfonts\fontsize\sf@size\z@\selectfont VI}%
}
\makeatletter
\DeclareRobustCommand{\CaVIII}{%
  \mbox{Ca \check@mathfonts\fontsize\sf@size\z@\selectfont VIII}%
}
\makeatother
\makeatletter
\DeclareRobustCommand{\CaIV}{%
  \mbox{Ca \check@mathfonts\fontsize\sf@size\z@\selectfont IV}%
}
\makeatletter
\DeclareRobustCommand{\MgIV}{%
  \mbox{Mg \check@mathfonts\fontsize\sf@size\z@\selectfont IV}%
}
\makeatother
\makeatletter
\DeclareRobustCommand{\MgVII}{%
  \mbox{Mg \check@mathfonts\fontsize\sf@size\z@\selectfont VII}%
}

\submitjournal{ApJ}

\shorttitle{Dual AGN in dwarf-dwarf galaxy mergers}
\shortauthors{Mićić et al.}
\graphicspath{{./}{figures/}}

\begin{document}

\title{Two Candidates for Dual AGN in Dwarf-Dwarf Galaxy Mergers}

\correspondingauthor{Marko Mićić}
\email{mmicic@crimson.ua.edu}
\author[0000-0003-3726-5611]{Marko Mićić}
\affiliation{Department of Physics \& Astronomy, University of Alabama, Tuscaloosa, AL 35401, USA}

\author[0000-0003-0100-0339]{Olivia J. Holmes}
\affiliation{Department of Physics \& Astronomy, University of Alabama, Tuscaloosa, AL 35401, USA}

\author[0000-0002-5087-8183]{Brenna N. Wells}
\affiliation{Department of Physics \& Astronomy, University of Alabama, Tuscaloosa, AL 35401, USA}

\author[0000-0003-4307-8521]{Jimmy A. Irwin}
\affiliation{Department of Physics \& Astronomy, University of Alabama, Tuscaloosa, AL 35401, USA}



\begin{abstract}

Dual AGN are important for understanding galaxy-merger-triggered fueling of black holes and hierarchical growth of structures. The least explored type of dual AGN are those associated with mergers of two dwarf galaxies. According to observations and cosmological simulations, dwarf galaxies are the most abundant type of galaxies in the early Universe and the galaxy merger rate is dominated by dwarfs. However, these mergers are generally too distant to be directly observed, and low-redshift dwarf-dwarf merger-related dual AGN are notoriously hard to find. In this paper, we present the first results of our large-scale search for this elusive type of object and the first two candidates for dual AGN in dwarf-dwarf mergers. Both objects exhibit tidal features (tails and bridges) characteristic of galaxy mergers/interactions. One object is apparently in a late-stage merger with an AGN separation of $<$ 5kpc, while the second is in an early-stage merger with interacting galaxies having established a tidal bridge. Both objects have dual, luminous X-ray sources that are most likely due to actively accreting massive black holes. Also, both objects have infrared counterparts, with colors consistent with being AGN. Follow-up observations will provide us a glimpse into key processes that govern the earliest phases of growth of galaxies, their central black holes, and merger-induced star formation. 

\end{abstract}



\section{Introduction} \label{sec:intro}
\begin{figure*}[t]
\epsscale{1.2}
\plotone{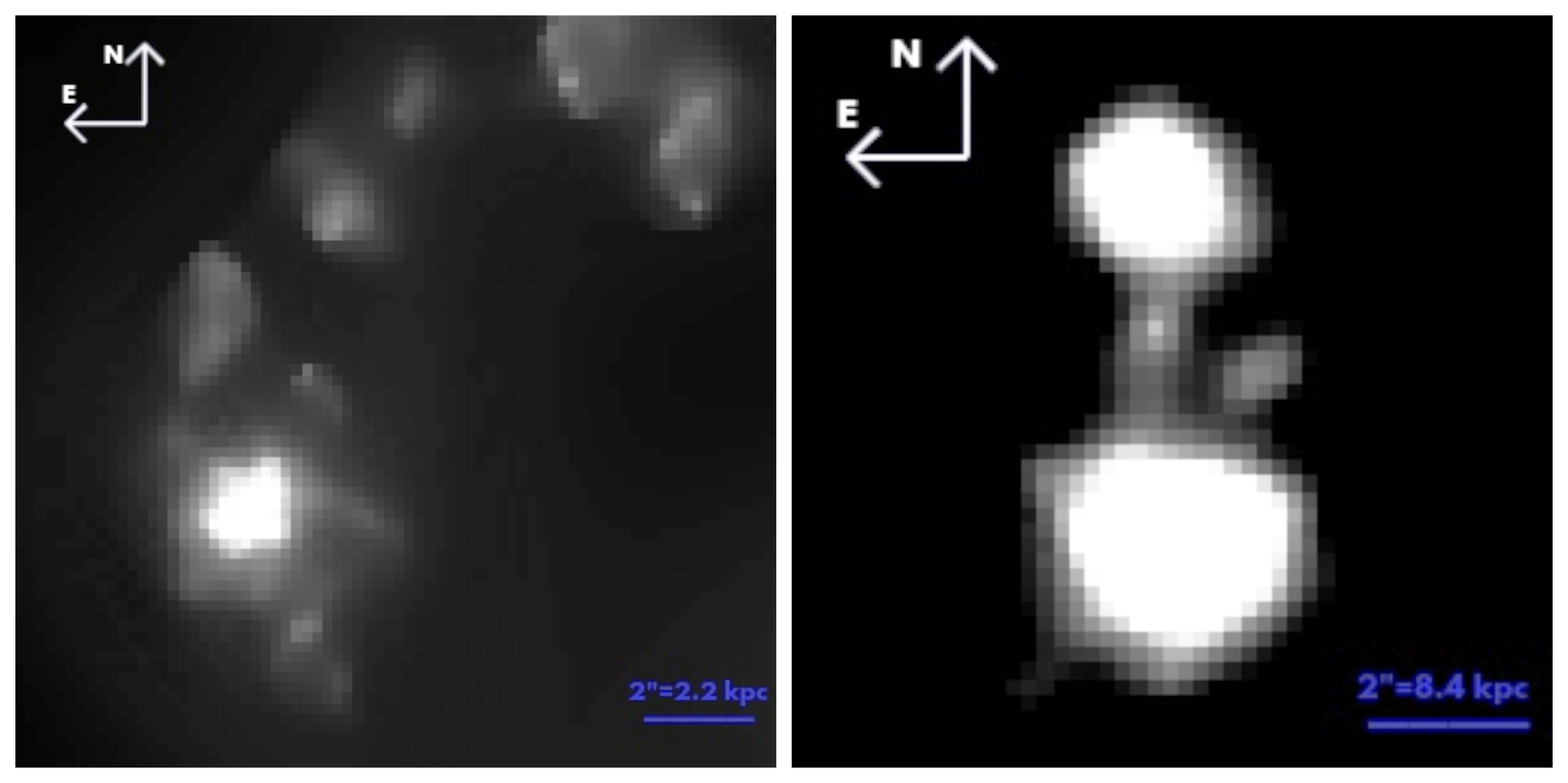}
\caption{Left panel: Adaptively smoothed CFHT \textit{r}-band image of Mirabilis. The galaxy is associated with a 15\arcsec  long tidal tail, and a smaller counter-tail, evidence of a late-stage galaxy merger. Right panel: Adaptively smoothed and contrast-enhanced CFHT \textit{r}-band image of Elstir and Vinteuil. Galaxies are connected with a tidal bridge, evidence of an early-stage galaxy merger.
\label{fig:fig1}}
\end{figure*}

Dual active galactic nuclei (DAGN) are systems consisting of two actively accreting massive black holes, typically separated by a few tens of kiloparsecs or less. Since every galaxy is expected to have a central massive black hole, DAGN are a natural consequence of galaxy mergers. DAGN can be significant for studies of merger-triggered accretion \citep{2015ApJ...806..219C}, hierarchical growth of structures, and gravitational waves \citep{2019ApJ...879L..21G}. Due to their importance for many fundamental astrophysical questions, DAGN have garnered significant attention in recent years. Many large-scale surveys have been carried out to uncover large populations of DAGN, using various approaches and methods, including X-rays \citep{10.1111/j.1365-2966.2008.13078.x}, optical \citep{Kim_2020}, near- and mid-infrared \citep{Satyapal_2017}, radio observations \citep{Rodriguez_2006}, or using combinations of optical and X-ray \citep{Teng_2012}, infrared and X-ray \citep{Pfeifle_2019}, etc. However, all these attempts have had only moderate success, and the number of known DAGN remains disproportionately small compared to the time and resources invested in these searches. One of the reasons for the low success rate is a bias towards major mergers of large galaxies, while dwarf-related mergers are neglected. For example, some of the aforementioned works require a preselected sample of galaxies with available spectra, or a sample of merging galaxies selected from SDSS or Galaxy Zoo \citep{10.1111/j.1365-2966.2008.13689.x}, which, except for the most nearby, eliminates the majority of faint, small, dwarf galaxies. \newline \indent
Dwarf galaxies are the most common type of galaxies, but there has not been a systematic survey for AGN in dwarf-related mergers. From the galaxy stellar mass-black hole mass relation it is expected for dwarf galaxies to harbor low-mass central black holes \citep{Reines_2015}. Some intermediate-mass black hole (IMBH) candidates have been identified using X-ray observations in the dwarf galaxies located on the outskirts of their massive galaxy companions, while undergoing active accreting phases, possibly triggered by an ongoing merger \citep{2012ApJ...747L..13F,10.1093/mnras/stt1459,Lin_2016,2017ApJ...836..183S}, demonstrating the importance of minor mergers. Dwarf-dwarf mergers are potentially the most interesting type of mergers since theoretical predictions and observations of ultraviolet luminosity functions suggest that dwarf galaxies are the most common type of galaxies at high redshifts \citep{Scannapieco_2002,Mason_2015} and that the galaxy merger rate is dominated by dwarfs. Also, according to the standard $\Lambda$CDM cosmological paradigm, the growth of structures happens hierarchically through mergers \citep{10.1111/j.1365-2966.2009.14396.x}. Thus, observations of dwarf-dwarf mergers, studies of their star-formation properties, and AGN content play an essential role in understanding the earliest stages of galaxy formation and evolution. However, these systems are very faint and impossible to observe directly at large redshifts, but detecting their low redshift analogs can give us a glimpse into processes that commonly happened in the early Universe but remained elusive. There have been works focused on discovering dwarf-dwarf mergers \citep{2015ApJ...805....2S,Paudel_2018}, but no systematic survey for DAGN in these objects has been done. The poster-child example for AGN in a dwarf-dwarf merger is Mrk 709 \citep{Reines_2014, Kimbro_2021}. Mrk 709 consists of two dwarf galaxies in the early stages of a merger, connected by a long, tidal bridge. Only one galaxy hosts an X-ray- and radio-detected AGN candidate, and both galaxies and the bridge are sites of intensive star formation. The absolute magnitudes of both galaxies are $\approx$--20, with stellar masses just below 3$\times$10$^{9}$ \(\textup{M}_\odot\), the mass usually used as an upper limit when defining dwarf galaxies \citep[e.g.][]{Reines_2020,Lemons_2015}. \cite{2011ApJ...742...68J} found that 9$\%$ of low-mass AGN from their sample reside in dwarf galaxies with a possible companion. Also, \cite{2018MNRAS.478.2576M} found that eight out of 23 AGN from their sample are associated with dwarf galaxies with a potential companion. However, it is unclear whether these galaxies are actually interacting or if they are chance aligned, and there is no evidence for the existence of DAGN in any of these systems. \newline \indent
\begin{deluxetable}{ccccc}
\tablenum{1}
\tablecaption{Summary of \textit{Chandra} ACIS-I observations\label{tab:tab1}}
\tablewidth{0pt}
\tablehead{\colhead{Galaxy} &\colhead{ID}& \colhead{Date}  & \colhead{t$_{exp}$} & \colhead{Off-axis angle}\\
\colhead{}&\colhead{}&\colhead{}&\colhead{(ks)}&\colhead{}\\
}
\startdata
Mirabilis & 13446 & 2011-09-09&58.42&5\farcm2\\
 & 13449 & 2011-06-06&68.15&2\farcm3\\
 & 14338 & 2011-09-10&117.5&5\farcm2\\ \hline
Elstir \& & 13997 & 2012-09-27&27.64&3\farcm7\\
Vinteuil &15538&2012-09-28&93.33&3\farcm7\\
&15540&2012-10-09&26.72&4\farcm5\\
\enddata
\end{deluxetable}
In this paper, we present the first two DAGN candidates in dwarf-dwarf mergers. The first DAGN candidate is associated with an Abell 133 dwarf galaxy, which is currently in a late-stage merger and exhibits a long, clumpy tidal tail. We name this galaxy \textit{Mirabilis} after an endangered species of hummingbird known for their exceptionally long tails. The second candidate is associated with two dwarf galaxies in an early-stage merger, connected with a faint tidal bridge associated with the Abell 1758S cluster. We name these two galaxies \textit{Elstir} and \textit{Vinteuil} after fictional artists from Marcel Proust's ``In Search of Lost Time''. Both objects are associated with dual, luminous X-ray sources, and infrared AGN-consistent counterparts and are shown in Figure \ref{fig:fig1}. The paper is organized as follows: In Section \ref{sec:data} we present the data used in this work and describe the ongoing large-scale project, and data analysis methods; in Section \ref{sec:res} we present properties of galaxies and their tidal structures, spectral analysis of X-ray sources, and analysis of infrared sources; in Section \ref{sec:dis} we discuss our results and alternate explanations.
\section{Methodology and Data Analysis} \label{sec:data}
This paper is a part of a broader survey in which we search for hidden DAGN in previously unknown, unrecognized galaxy mergers, systems that have been missed by all previous works. In order to assess AGN candidacy, we use X-ray and infrared observations. We surveyed the \textit{Chandra} archival database and found that $\approx$ 20 deg$^2$ of the sky has been observed with at least 100 ksec of exposure time, often in multiple, overlapping observations. Once these overlapping, identical observations are collected and stacked, deep X-ray images can be created, allowing for many faint, previously unknown, and uncatalogued X-ray sources to be revealed. The X-ray approach is one of the most fruitful methods used to detect AGN since accreting black holes are expected to produce substantial X-ray emission. Also, our methodology in which we only utilize deep, stacked X-ray data will allow detection of faint AGN, usually associated with dwarf galaxies. An increasing number of X-ray detected AGN in dwarf galaxies have been discovered recently \citep{Lemons_2015,Pardo_2016,10.1093/mnras/staa040} and many of them have been observed to have X-ray luminosities $<$10$^{41}$ erg s$^{-1}$, far below what is usually adopted for AGN X-ray luminosity in regular galaxies. Dwarf galaxies are more likely to contain lower-mass black holes, which are expected to have lower intrinsic X-ray luminosities \citep{simmrefId0}. Also, AGN in galaxy mergers are likely to be heavily obscured, particularly in late-stage mergers \citep{Pfeifle_2019}, making intrinsically faint X-ray sources even fainter. Therefore, large exposure times are required to detect these AGN, even at modest distances. We mitigate this issue  by only using stacked, deep X-ray data with lower detection limits. The sensitivity throughout our data sets varies, but in general X-ray sources as faint as 10$^{39}$ erg s$^{-1}$ can be detected. The first step in our methodology is collecting and stacking X-ray data to create deep X-ray images. Then, we use the \textbf{CIAO} tool \textbf{wavdetect} with wavelets 1.4, 2, 4, 8, to detect all sources in the X-ray image. We then collect these sources, calculate distances between every possible pair, and keep only those separated by no more than 10\arcsec.\newline\indent
We then proceed to the second step, in which we crossmatch potential dual X-ray sources with a catalog of infrared detected AGN-consistent objects, compiled with the help of WISE telescope. Galaxies that are red in their WISE W1--W2 ([3.4$\upmu$m]--[4.6$\upmu$m]) color contain heated dust, and the reddest systems are classified as AGN \citep{10.1093/mnras/stw1976}. The lower limit on W1--W2 for AGN varies from 0.5 to 0.8, depending on the study. We used a catalog of AGN candidates, derived from WISE and based on the aforementioned criterion that contains more than 20 million sources \citep{2018ApJS..234...23A}. Since \textit{Chandra} has far superior resolution than WISE, we only require the presence of one AGN-consistent infrared counterpart since, in general, WISE would not be able to distinguish sources separated by less than 5\arcsec. We introduce this step because in some cases, especially with low luminosity X-ray sources, X-ray detection alone would not be enough to claim the AGN nature of the source, and WISE provides additional evidence for the AGN scenario. Upon completing $\approx$20\% of X-ray data, and after applying the first two steps of our methodology, we ended up with 497 potential dual X-ray sources with at least one AGN-consistent IR counterpart.\newline \indent
In the third step we try to establish if the potential dual X-ray+IR sources are chance-aligned objects or genuine pairs. One of the ways the physical separation can be calculated is if the host galaxies have available spectroscopic redshifts. However, most galaxies, especially the faint ones, do not have readily available spectra. Another way is to detect tidal features that usually form during galaxy mergers or interactions, such as tidal tails, arms, bridges or shells. Such tidal features would be a clear evidence of galaxy association, which would imply that the dual X-ray+IR sources are bona fide pairs. For this purpose, we use the archival data from the Canadian-French-Hawaiian Telescope (CFHT) since in most occasions the CFHT had available archival data. However, the tidal features are often very faint and hard to notice. Therefore, we apply careful image processing to the CFHT observations. Specifically, we use an adaptive smoothing procedure in which the number of counts under each kernel is set to a certain value. Upon applying adaptive smoothing on optical CFHT images of 452 potential DAGN, we detect tidal debris in three cases. In two out of those three cases, the properties of merging galaxies, derived from the available data, indicated that the most likely explanation is a dwarf-dwarf merger. The third case is most likely a minor merger between a $z=$0.9 large galaxy and its dwarf satellite, and as such is not included in this paper. The other 449 potential DAGN did not show any signs of interaction-produced tidal debris, and are likely unrelated, chance-aligned objects. Even though the percent of bona fide galaxy mergers is small, adaptive smoothing turned out to be quite powerful tool since tidal debris detected in three cases is not visible in SDSS, Galaxy Zoo, or any other large scale survey due to its faintness. Tidal debris is only revealed upon applying adaptive smoothing and would have stayed undetected otherwise. Also, the galaxy Vinteuil, shown in right panel of the Figure \ref{fig:fig1}, does not exist in any catalog, and was catalogued for the first time in this project, thanks to the adaptive smoothing procedure. This demonstrates the power and importance of the image processing technique described in this paper since it can significantly enhance the quality of average optical observations, allowing to reveal the structures of extremely low surface brightness.
\newline \indent
The X-ray data reduction was performed using \textbf{CIAO 4.12} tools, and Calibration Database \textbf{CALDB 4.9.3}. All observations were reprocessed using the \textbf{chandra\_repro} tool. Then, using \textbf{merge\_obs} tool observations are reprojected and combined to create a merged event file and exposure-corrected images. Optical image processing was performed using the \textbf{CIAO 4.12} adaptive\_smoothing tool. Data sets with potential DAGN were corrected for astrometry. We start by running the CIAO tool wavdetect across the full field of view of X-ray image in order to detect X-ray sources. The parameters were set in such way to focus only on strong sources with robust centroids. Then we matched \textit{Chandra} sources with the USNO-A2.0 catalog. The CIAO tool wcs$\_$match takes in previously selected X-ray and USNO-A2.0 sources, iteratively crossmatches them, and determines transformation parameters to minimize positional differences. Next, the CIAO tool wcs$\_$update is used to apply the correction parameters and update the \textit{Chandra} event and aspect solution files.
\begin{figure*}[t]
\epsscale{1.2}
\plotone{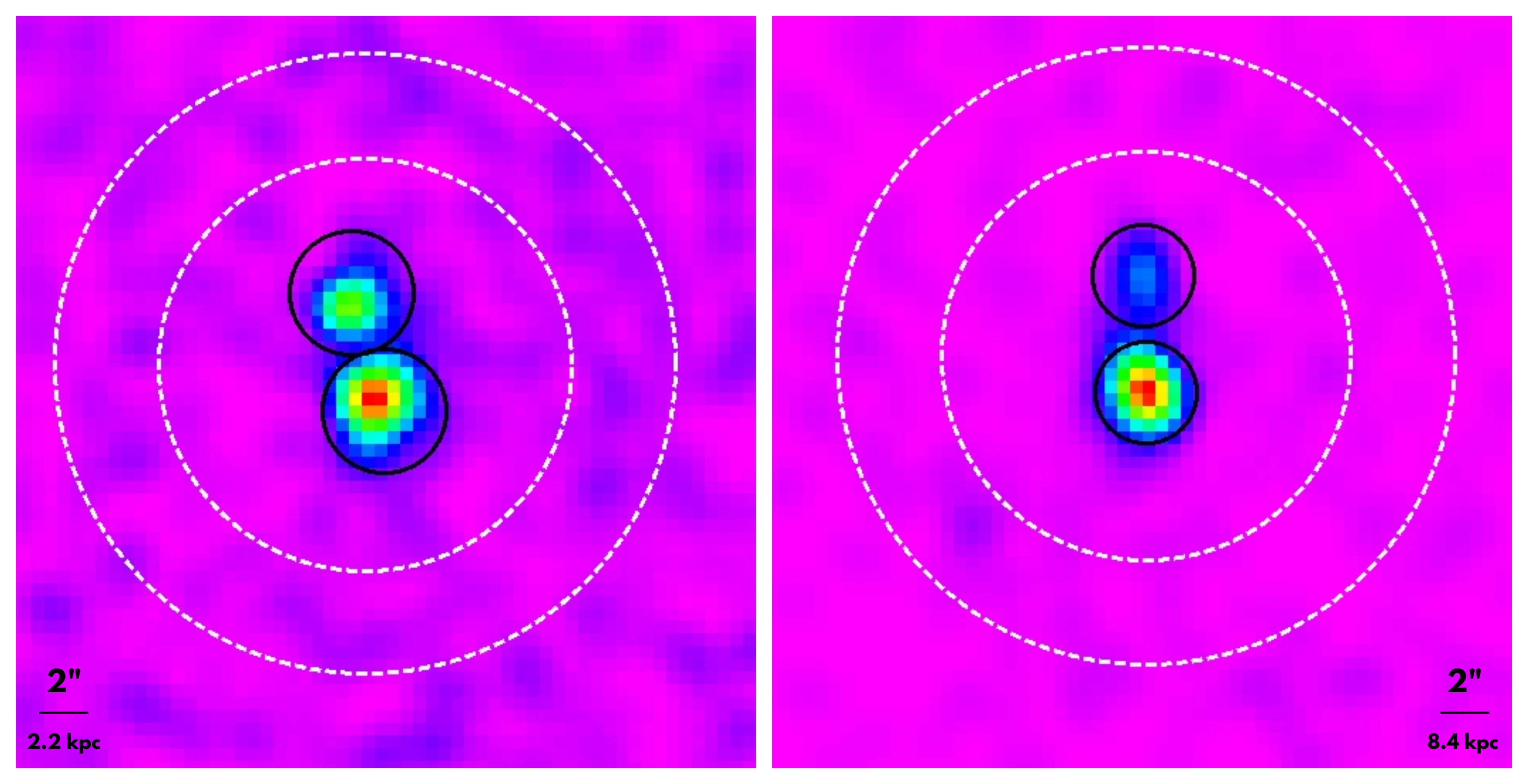}
\caption{Left: Stacked \textit{Chandra} X-ray image of Mirabilis showing X1 (southern) and X2 (northern) source. Right: Stacked \textit{Chandra} X-ray image of Elstir and Vinteuil showing X3 (southern) and X4 (northern) source. Black solid circles represent source extraction regions (80\% time exposure-weighted point spread function). White dashed annular regions represent background regions used to calculate net counts for each source.
\label{fig:fig2}}
\end{figure*}

\begin{figure*}[t]
\epsscale{1.2}
\plotone{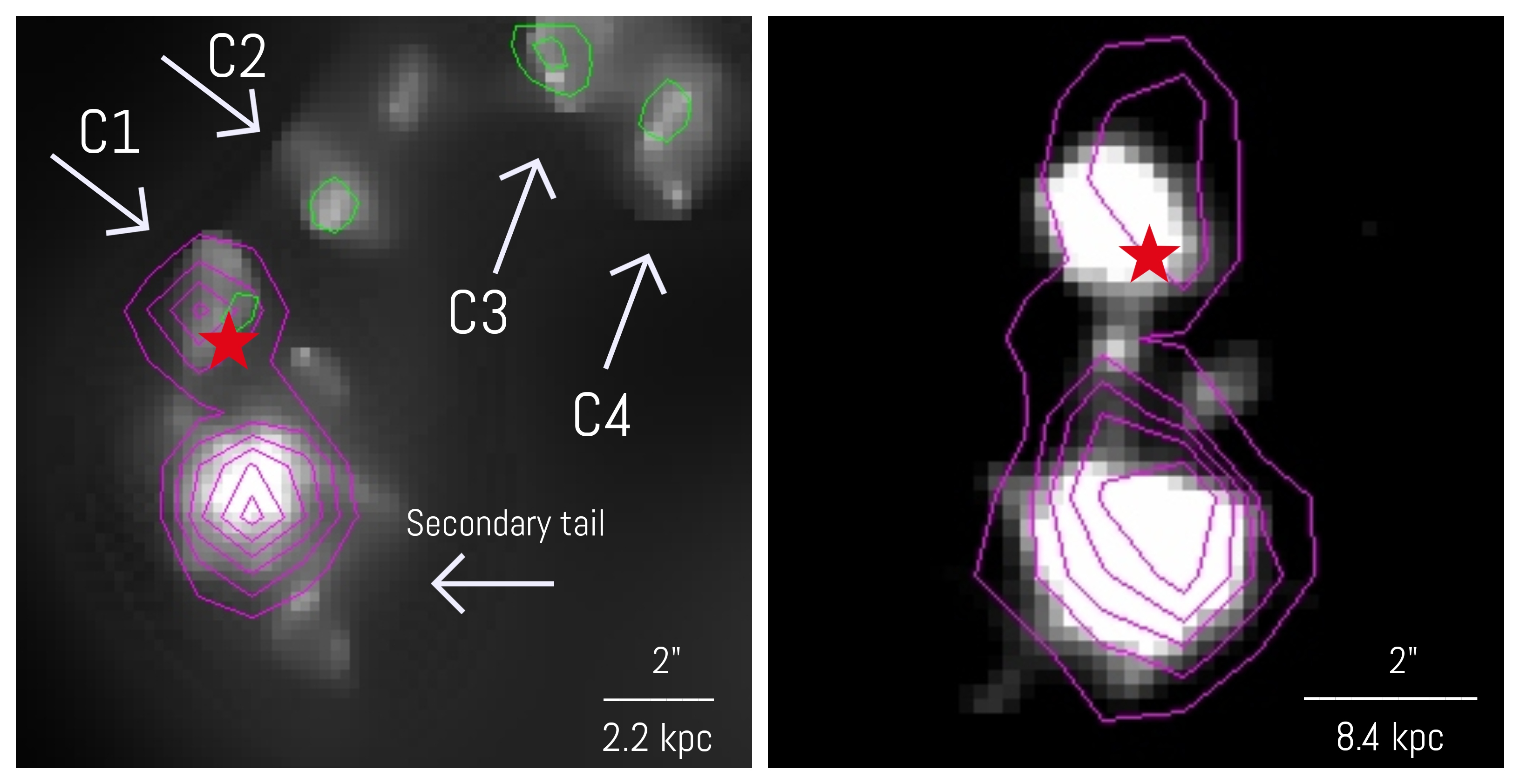}
\caption{Left: CFHT \textit{r}-band image of Mirabilis. Overlaid magenta-colored contours represent positions of \textit{Chandra} detected X-ray sources. The southern source spatially coincides with the main galaxy, while the northern source coincides with the first clump of the tail. Four clumps labeled with C1 through C4 were detected in Dark Energy Survey. Green contours represent CFHT \textit{g}-band detected clumps. Green contours for the main galaxy are omitted for clarity. CFHT \textit{g}-band detected clumps match with CFHT \textit{r}-band and Dark Energy Survey detected clumps. The red star symbol marks the location of a WISE-detected infrared AGN-consistent source. A secondary counter-tail can be seen extending from the opposite side of the main galaxy, forming a rough S shape with the main tail. Right: Adaptively smoothed CFHT \textit{r}-band image of Elstir and Vinteuil. Magenta-colored contours represent positions of \textit{Chandra} detected X-ray sources. The red star symbol marks the location of a WISE-detected infrared AGN-consistent source.
\label{fig:fig3}}
\end{figure*}
\section{Results} \label{sec:res}
\subsection{Properties of Host Galaxies}
Mirabilis, shown in the left panel of Figure \ref{fig:fig1}, is classified as a member of the Abell 133 galaxy cluster \citep{refId0}, located at the edge of the cluster. Adopting the redshift of Abell 133, $z$=0.056338 \citep{2004AJ....128.1558S}, to be Mirabilis' distance, its \textit{g}- and \textit{r}-band absolute magnitudes become --13.9 and --14.2, respectively. At this distance 1$\arcsec$ corresponds to $\approx$ 1.1 kpc, and Mirabilis' Kron radius is $\approx$ 3.5 kpc \citep{refId0}.\newline \indent
Elstir, the southern galaxy of the system shown in the right panel of Figure \ref{fig:fig1}, was cataloged in SDSS data release 16 \citep{2020ApJS..249....3A}, and it is cross-identified as SDSS J133300.32+502332.1. Its photometric redshift is 0.22$\pm$0.13 \citep{2009ApJS..182..543A}. Also, Elstir is superposed onto galaxy cluster Abell 1758S whose spectroscopic redshift is 0.2729 \citep{10.1111/j.1365-2966.2009.14823.x}, and we adopt this value as Elstir's redshift. At this redshift 1$\arcsec$ corresponds to 4.2 kpc. Elstir, just like Mirabilis, is located at the very edge of its host cluster. Elstir's \textit{g}- and \textit{r}-band absolute magnitudes are --17.1 and --18.7. It is significantly redder than Mirabilis, indicating its larger distance. Its Petrosian radius is 1\farcs55, or 6.5 kpc. Vinteuil, the northern galaxy of the second system, has not been catalogued and no information on its magnitudes is available. However, since optical observations revealed a tidal bridge connecting it with Elstir we can safely adopt the same redshift of $z$=0.2729. When compared to its first neighbor, Vinteuil is noticeably smaller and fainter. 
\begin{figure*}[t]
\epsscale{1.2}
\plotone{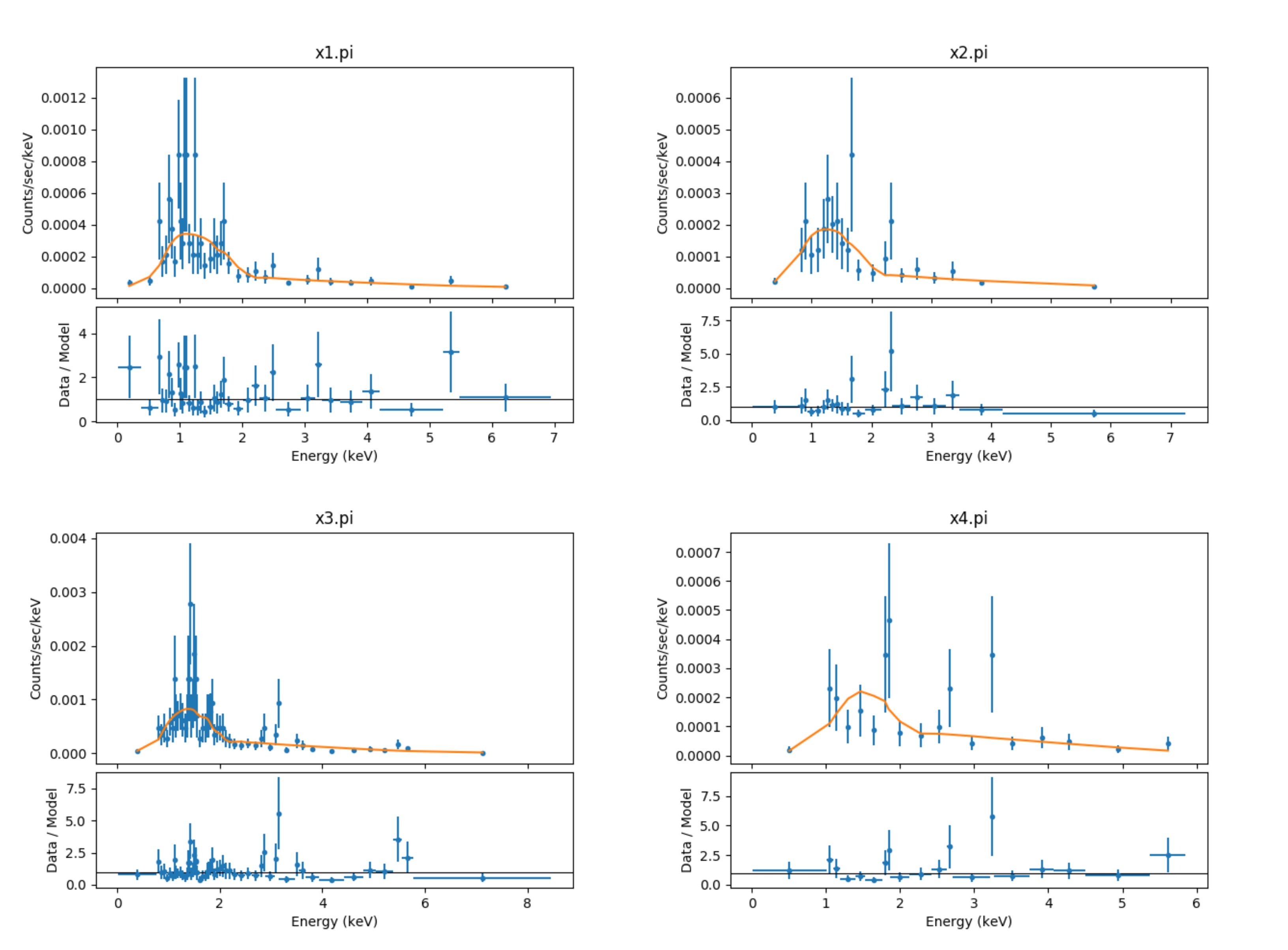}
\caption{Summary of best fit X-ray spectra for X1-X4.
\label{fig:fig4}}
\end{figure*}
\subsection{Optical observations}
\begin{deluxetable*}{cccccccccc}
\tablenum{2}
\tablecaption{Summary of properties of Mirabilis, Elstir and Vinteuil\label{tab:tab3}}
\tablewidth{0pt}
\tablehead{\colhead{Name} & \colhead{RA}  & \colhead{Dec} & \colhead{m$_g$}&\colhead{m$_r$} & \colhead{log L$_{0.5-8}$} &\colhead{log L$_{0.5-2}$}&\colhead{$\Delta$r}&\colhead{W1}&\colhead{W2}
}
\startdata
Mirabilis & 01:01:36.66 & --21:37:48.49 & 23.09 & 	22.83& 40.77&40.43 &0.7 &-- & --\\
C1 & 01:01:36.70 & --21:37:45.05 & 25.28 & 	25.29 & 40.55 &40.23&0.5& 16.84&15.99 \\
C2& 01:01:36.56 & --21:37:43.16 & 25.53 & 	25.57 & -- &--&--& --&--\\
C3 & 01:01:36.26 & --21:37:40.38 & 25.45 & 	25.19 & --&-- &--& --&-- \\
C4 & 01:01:36.12 & --21:37:41.30 & 25.79 & 	25.49 & --&-- &--& --&-- \\
Elstir & 13:33:00.32 & +50:23:32.19 & 23.65	 &	22.02 & 42.71 &42.37&0.4& --&-- \\
Vinteuil &13:33:00.38 & +50:23:36.35 & -- & -- & 42.29&41.96&0.2& 17.72&	16.23\\
\enddata
\tablecomments{m$_g$ and m$_r$ are \textit{g}- and \textit{r}-band apparent magnitudes \citep{2021ApJS..255...20A, 2020ApJS..249....3A}; log L$_{0.5-8}$ is logarithm of X-ray luminosity in the 0.5-8 keV band; log L$_{0.5-2}$ is logarithm of X-ray luminosity in the 0.5-2 keV band; $\Delta$r is angular separation in arcseconds between optical and X-ray sources (positions of X-ray sources are taken from \cite{2020AAS...23515405E}); W1 is [3.4$\upmu$m] WISE magnitude; W2 is [4.6$\upmu$m] WISE magnitude.}
\end{deluxetable*}
Interacting galaxies are likely to develop a variety of tidal features, such as arms, tails, bridges, and shells. For both objects we analyzed archival, $r-$band CFHT images, searching for indicators of ongoing merger. The adaptively smoothed image of Mirabilis revealed a $\approx$ 15$\arcsec$ long, clumpy tidal tail, and such tails usually form during galaxy interactions. In order to eliminate the possibility of the tail being an artifact of smoothing, we analyzed another CFHT observation, now in \textit{g}-band, and adaptive smoothing revealed the clumps at the same positions as \textit{r}-band clumps. In Figure \ref{fig:fig3} green contours represent CFHT \textit{g}-band detected contours overlaid on CFHT \textit{r}-band image. Another possibility is that the clumps are serendipitously aligned background galaxies, creating an illusion of a tail. Four of the brightest clumps, labeled as C1 through C4 in Figure \ref{fig:fig3}, were detected in the Dark Energy Survey (DES) \citep{2021ApJS..255...20A}, and their \textit{g}-, \textit{r}-band magnitudes, and \textit{g}--\textit{r} colors are all within $\pm$0.2, i.e., within the DES magnitude uncertainty. Assuming the clumps are background galaxies, they would have various uncorrelated intrinsic luminosities and distances, and it is very unlikely that the distribution of these parameters is such to result in nearly identical magnitudes and colors. On the other hand, the observed properties are expected if the clumps are arising from a homogeneous structure, such as a tidal tail. Another form of evidence that the tail and merger are real is a small, secondary tail extending from Mirabilis, on the opposite (southern) side from the main tail. If the masses of the galaxies are comparable, two tidal tails are expected to form, usually forming a characteristic S shape \citep{10.1093/mnras/staa2985}, which is what we observe in the case of Mirabilis. We interpret the larger tail to be a remnant of a galaxy. The merger is likely to be in its late stages, i.e., multiple passages happened, and the smaller galaxy was completely tidally disrupted by its more massive companion.
\begin{deluxetable}{ccccc}
\tablenum{3}
\tablecaption{Summary of the X-ray spectral fitting properties\label{tab:tab2}}
\tablewidth{0pt}
\tablehead{\colhead{ID} & \colhead{Cstat/d.o.f.}  & \colhead{N$_{\textup{H}}$} & \colhead{F$_\textup{{0.5-8}}$}&\colhead{F$_\textup{{0.5-2}}$}\\
}
\startdata
X1 & 38.4/38 & $<$0.1 &7.6$^{+0.3}_{-0.2}$ &3.5$^{+0.5}_{-0.3}$\\
X2 & 18.4/20 & 0.1$\pm$0.1 & 4.6$^{+0.3}_{-0.6}$  &2.2$^{+0.5}_{-0.7}$\\
X3 & 55.1/56 & 0.5$\pm$0.1 & 22$^{+1}_{-3}$ &10$^{+1}_{-1}$\\
X4 & 23.6/18 & 1.3$\pm$0.4 & 8.2$^{+1.6}_{-3.6}$ &3.9$^{+0.4}_{-1.1}$\\
\enddata
\tablecomments{Cstat/d.o.f. is the Cstatistics of the best fit and number of degrees of freedom, N$_{\textup{H}}$ is intrinsic absorption in units of 10$^{22}$ cm$^{-2}$, F$_\textup{{0.5-8}}$ is the measured unabsorbed flux in the 0.5-8 keV band in units of $\times$10$^{-15}$ erg s$^{-1}$ cm$^{-2}$, F$_\textup{{0.5-8}}$ is the measured unabsorbed flux in the 0.5-8 keV band in units of $\times$10$^{-15}$ erg s$^{-1}$ cm$^{-2}$. $\Gamma$ was kept fixed at 1.9. The listed uncertainties are estimated at 1$\sigma$ confidence level.}
\end{deluxetable}\newline\indent
Similar adaptive smoothing applied on another archival CFHT $r-$band image containing Elstir and Vinteuil revealed a faint tidal  bridge connecting the two, indicating an ongoing interaction. The galaxies are likely in the early stages of the merger, i.e., in their first passage, and Elstir is stripping material off Vinteuil, creating the tidal bridge.
\subsection{X-Ray Observations}
Mirabilis was observed with the \textit{Chandra} ACIS-I instrument on three occasions in 2011 (PI: Vikhlinin), with a total exposure time of $\approx$ 244 ksec. Stacked \textit{Chandra} image revealed a dual X-ray source. \textit{Chandra} X-ray magenta-colored contours are overlaid in Figure \ref{fig:fig3}. It can be seen that the southern source (X1 hereafter) coincides with the position of Mirabilis, while the northern source (X2 hereafter) coincides with the first clump of the tail (C1). The sources are separated by $\approx$4\arcsec, or $\approx$4.4 kpc. The source extraction regions were circular apertures, with the radius corresponding to the time exposure weighted 80\% point spread function ($\approx$ 2\farcs6). The background region was an annular region, with inner and outer radii of 8\arcsec and 12\arcsec. The derived net counts of X1 and X2 were 125$\pm$12 and 65$\pm$9 counts, respectively.\newline \indent
Elstir and Vinteuil were serendipitously observed with the \textit{Chandra} ACIS-I instrument on three occasions in 2012 (PI: David), with a total exposure time of $\approx$ 148 ksec. The stacked \textit{Chandra} image revealed a dual X-ray source. The magenta-colored contours are overlaid on the CFHT \textit{r}-band image in Figure \ref{fig:fig3}. The southern X-ray source spatially coincides with Elstir (X3 hereafter) and the northern X-ray source with Vinteuil (X4 hereafter). The sources are separated by $\approx$ 5$\arcsec$, or $\approx$ 21 kpc. The source extraction regions were circular apertures with a radius of $\approx$ 2$\arcsec$, derived in the same way as in the case of Mirabilis, and the background region was an annular region, with inner and outer radii of 8\arcsec and 12\arcsec. The net counts of X3 and X4 were 194$\pm$14 and 56$\pm$8 counts, respectively.\newline \indent
We calculated offsets between optical and X-ray sources and reported them in Table \ref{tab:tab3}. The offsets range from 0\farcs2 to 0\farcs7. Then, we calculated the count contribution weighted positional uncertainties, at 95\% confidence level, following equation 12 from \cite{2007ApJS..169..401K}. We found that in all four X-ray sources the positional uncertainties were larger than offsets reported in Table \ref{tab:tab3}.
\subsection{X-Ray Spectral Fitting}
In  order  to  translate  count  rate  to  flux, we used \textbf{SHERPA}, CIAO’s modeling and fitting application. First, we used the CIAO tool \textbf{specextract} to obtain the instrumental response functions, both the auxiliary response file (ARF) and the redistribution matrix file (RMF) for each observation. Then, we used the \textbf{combine\_spectra} tool to create summed source counts, response, and background files. Since our sources have a low number of counts, we used Cstat statistics to perform all the fits since it does not require counts binning to work and is usually adopted for low-counts spectral fitting. We specified a source model composed of two components: a photoelectric absorption model and a power-law emission model, \textbf{xszphabs*powlaw1d}. The fit is performed by fixing the photon index to be $\Gamma$=1.9, a
typical AGN value, and leaving the intrinsic absorption
N$_{\textup{H}}$ free to vary. Then, the fit is repeated leaving both
$\Gamma$ and N$_{\textup{H}}$ free to vary, and we check if the fit is significantly improved, (i.e., if Cstatold–Cstatnew $>$ 2.71,
where Cstatold and Cstatnew are the Cstat values for
the fit without and with $\Gamma$ free to vary, respectively). This procedure is adopted from \cite{Marchesi_2016} and \cite{2018MNRAS.478.2576M}, and we refer the reader to these papers for more details. All four X-ray sources showed better fit results with photon index fixed at $\Gamma$=1.9. The resulting X-ray luminosities for X1, X2, X3 and X4 were 5.9 $\times$ 10$^{40}$, 3.5 $\times$ 10$^{40}$, 5.1 $\times$ 10$^{42}$, and 2.0 $\times$ 10$^{42}$ erg s$^{-1}$, respectively. The procedure was repeated with photon indices fixed at 1.7 and 2.1, but the luminosities did not significantly differ from $\Gamma$=1.9 values. The summary of the X-ray spectral fitting procedure is shown in Table \ref{tab:tab2}. The best fits for all four sources are shown in Figure \ref{fig:fig4}.
\begin{figure*}[t]
\epsscale{1.2}
\plotone{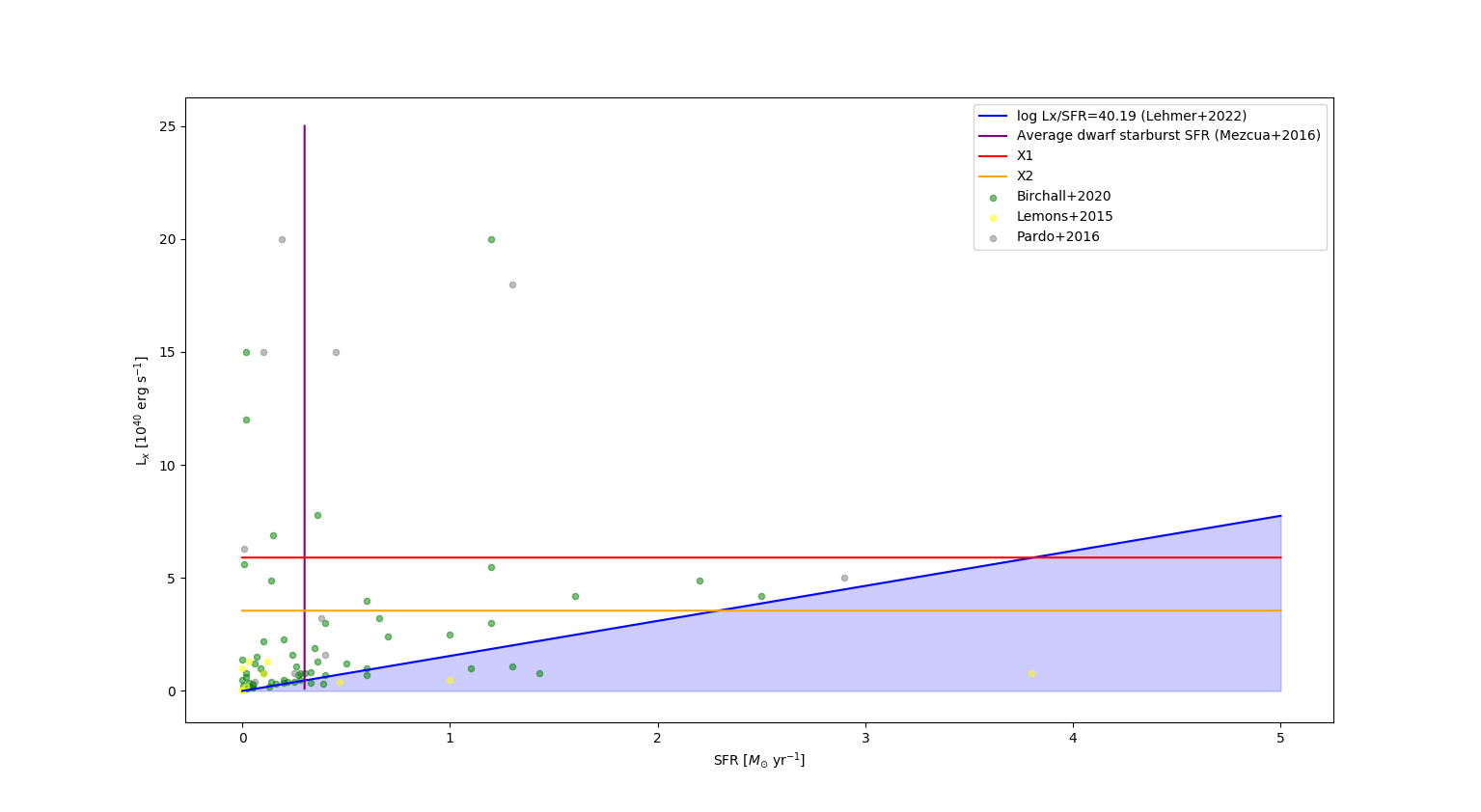}
\caption{The blue diagonal line shows the elevated relation between SFR and HMXB produced X-ray luminosity \citep{2022ApJ...930..135L}. Vertical purpleline shows average SFR of local dwarf starburst galaxies \citep{Mezcua_2016}. Horizontal red and orange lines show luminosities of our sources X1 and X2 with unknown SFR. Their intersection with the diagonal blue line shows required SFR for observed X-ray luminosity to be due to populations of HMXBs. Green, yellow and grey points show X-ray luminosities and SFRs of dwarf galaxies from various papers \citep{Pardo_2016,Lemons_2015,10.1093/mnras/staa040}. Majority of the AGN candidates in dwarf galaxies from the literature are above the blue diagonal line, thus still viable AGN candidates even when an elevated relation is used to estimate X-ray emission from stellar processes.
\label{fig:fig6}}
\end{figure*}
\subsection{Infrared Observations}
The X2 source has an infrared counterpart detected with WISE, with W1--W2 color consistent with being an AGN, and it was cataloged as an AGN candidate \citep{2018ApJS..234...23A}. The X1 source does not have an infrared counterpart, however, the spatial resolution of WISE is such that even if there was a second infrared source they could not be spatially separated. The position of the infrared source is shown with a red star symbol in Figure \ref{fig:fig3}. \newline \indent
Similarly, the X4 source has an infrared WISE-detected counterpart, classified as an AGN candidate, due to its W1--W2 color. The position of the infrared source is labeled with a red star symbol in Figure \ref{fig:fig3}. The same as in the case of Mirabilis, WISE's spatial resolution is insufficient to distinguish sources separated by only 5$\arcsec$, thus the absence of the second infrared counterpart is expected.\newline \indent
\cite{10.1093/mnras/stw1976} noticed that the reddest infrared galaxies are a rising fraction of the low stellar mass galaxy population, even though many of these galaxies have not been optically classified as AGN or composite galaxies. \cite{Hainline_2016} studied 18,000 dwarf galaxies at redshifts $z<$0.055 using SDSS, WISE data and BPT-classification \citep{1981PASP...93....5B}, a diagnostics method which uses emission lines ratios to identify the primary ionization source in a galaxy. They found that some, but not all, BPT-classified AGN dwarf galaxies do have AGN-consistent infrared colors, but they also found that many BPT-classified starburst dwarf galaxies can produce similar AGN-consistent infrared colors. They concluded that star formation in dwarf galaxies can heat up the dust and mimic the infrared colors expected to emanate from AGN. However, \cite{Hainline_2016} relied on BPT classification axiomatically, but some recent works demonstrated that the accuracy of BPT diagnostics is a function of galaxy mass; BPT diagnostics are more accurate when working with the most massive galaxies with 93\% agreement, and least reliable when applied on the least massive galaxies with 30\% agreement \citep{10.1093/mnras/staa040}. Also, \cite{Schutte2022} observed dwarf galaxy Henize 2-10, one of the first known dwarf galaxies with an AGN candidate, and found that black hole driven outflows are triggering star formation. It becomes clear that one should be cautious when working with dwarf galaxies and a single indicator should not be used prove or refute the existence of AGN or star formation. Also, \cite{Hainline_2016} suggests that infrared colors are not a reliable AGN indicator when used alone, and both our infrared sources have luminous X-ray counterparts, that are likely to be due to AGN activity (see Subsection \ref{subsec:sub}).
\begin{figure}[t]
\epsscale{1.2}
\plotone{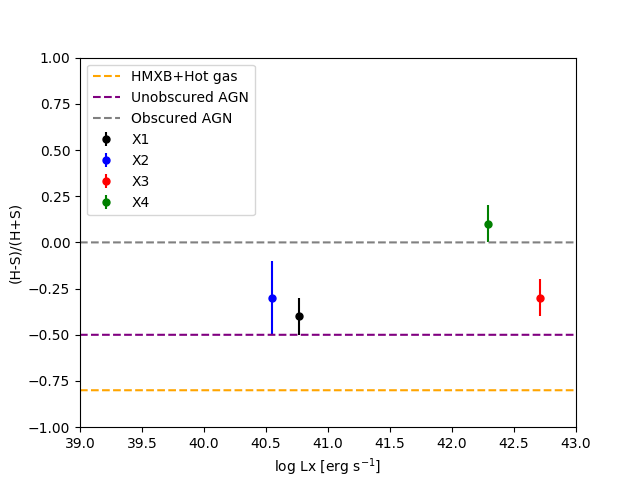}
\caption{Hardness ratio of four of our sources. Dashed horizontal lines show expected hardness ratios for X-ray emission produced by obscured AGN, unobscured AGN and combination of HMXBs and hot gas \citep{Pardo_2016}. Three sources reside in the unobscured AGN region, while the fourth one is located in the obscured AGN region.
\label{fig:fig7}}
\end{figure}
\section{Discussion \label{sec:dis}}
\subsection{Distance and Dwarf Nature of \textit{Mirabilis}}
Mirabilis was classified as a member of the Abell 133 galaxy cluster and adopting the redshift of the cluster to be its distance results in physical properties being in line with a dwarf nature. Mirabilis' physical properties (M$_r$=--14.2, r=3.5 kpc) imply it is a small, faint, low-mass dwarf galaxy. Adaptive smoothing revealed a long clumpy tidal tail which we interpret as evidence for an ongoing late-stage merger. The second galaxy was completely disrupted by Mirabilis' gravitational influence, suggesting it had to be significantly smaller than Mirabilis. The fact that Mirabilis' properties indicate it is an extremely small and faint dwarf, make this system a strong candidate for a dwarf-dwarf galaxy merger. The first clump of the tail, C1, hosts a luminous X-ray source and an infrared AGN-consistent source, hinting that the first clump is a remnant of the nucleus of a destroyed small, dwarf galaxy, while the rest of the tail is made up of the remaining, stripped material. \newline \indent Even though Mirabilis was classified as a member of the Abell 133 cluster \citep{refId0}, we do not have information about its photometric or spectroscopic redshift and it could be a background galaxy superposed onto Abell 133. In this case, its luminosity, size and stellar mass would increase, and if sufficiently distant, make it a regular galaxy instead of a dwarf. However, assuming the Abell 133 membership makes Mirabilis a very small dwarf. This implies Mirabilis would have to be considerably more distant than initially expected in order not to be a dwarf galaxy. We compare this scenario to the case of the previously mentioned Mrk 709 system, the merger of two borderline dwarf galaxies with evidence for only one AGN. Both galaxies in Mrk 709 have absolute magnitudes of $\approx$ --20 and $g-r$ colors comparable to that of Mirabilis. For Mirabilis to have an absolute magnitude of $\approx$--20 it would have to be at a redshift of $z$=0.7, and at this distance, Mirabilis would still likely be a Mrk 709-like borderline dwarf galaxy. In this scenario, the tail length would exceed 100 kpc, which would make it one of the longest tidal tails ever discovered. Placing Mirabilis at an even larger distance would make it a regular, non-dwarf galaxy, however, the whole system would grow even more in size, and the disrupted galaxy would still likely be within the dwarf regime since it had to be significantly smaller than Mirabilis. We conclude that Mirabilis represents a strong candidate for the first ever DAGN in a dwarf-dwarf merger. Even in the case of Mirabilis being a very distant background object would make this system a single dwarf galaxy merger, with DAGN, and record-breaking long tidal tail, a discovery interesting in its own right.
\subsection{The Dwarf Nature of Elstir and Vinteuil}
Unlike Mirabilis, Elstir has a photometric redshift that is, within the uncertainty, in line with the spectroscopic redshift of the cluster it is superposed onto, i.e., a strong indicator of Elstir's association with Abell 1758S cluster. Its Petrosian radius is $\approx$ 6.5kpc. \cite{zib10.1111/j.1365-2966.2004.07235.x} studied SDSS galaxies and found that the median $i$-band Petrosian radius is 10.6 kpc (assuming Hubble parameter of h=0.7). Their sample consisted of 1047 galaxies that had successfully measured $g$, $r$ and $i$ Petrosian magnitudes (Pmag), Pmag$_i>$17.5, $i-$band isophotal semi-major axis $a>$10\arcsec, and isophotal axis ratio $b$/$a\geq$ 0.25 According to this metric, Elstir is smaller than an average galaxy. Elstir's $g$-band absolute magnitude is $\approx$--17, and galaxies with absolute magnitudes fainter than --18 are considered to be dwarfs by many \citep[e.g.][]{2000AJ....119.2757V}. Also, the brightest dwarfs appear around absolute magnitude --18, while the faintest non-dwarf galaxies appear around absolute magnitude --16. Elstir is $\approx$15 times fainter than previously mentioned Mrk 709, and its $g$-band absolute magnitude is in line with local dwarfs, such as LMC or SMC. This information, together with a lower than median size by the Petrosian metric, strongly hints that Elstir is a dwarf galaxy. \newline \indent
We do not have information on Vinteuil's redshift or magnitudes, but since it is connected to Elstir by a tidal bridge, the same distance can be adopted and the two can be safely compared to each other. Vinteuil is noticeably smaller and fainter than Elstir, thus its dwarf nature is even more emphasized than in the case of Elstir. Therefore, we conclude that the Elstir-Vinteuil system represents a strong candidate for the second ever DAGN candidate in a dwarf-dwarf galaxy merger.
\begin{figure*}[t]
\epsscale{1.2}
\plotone{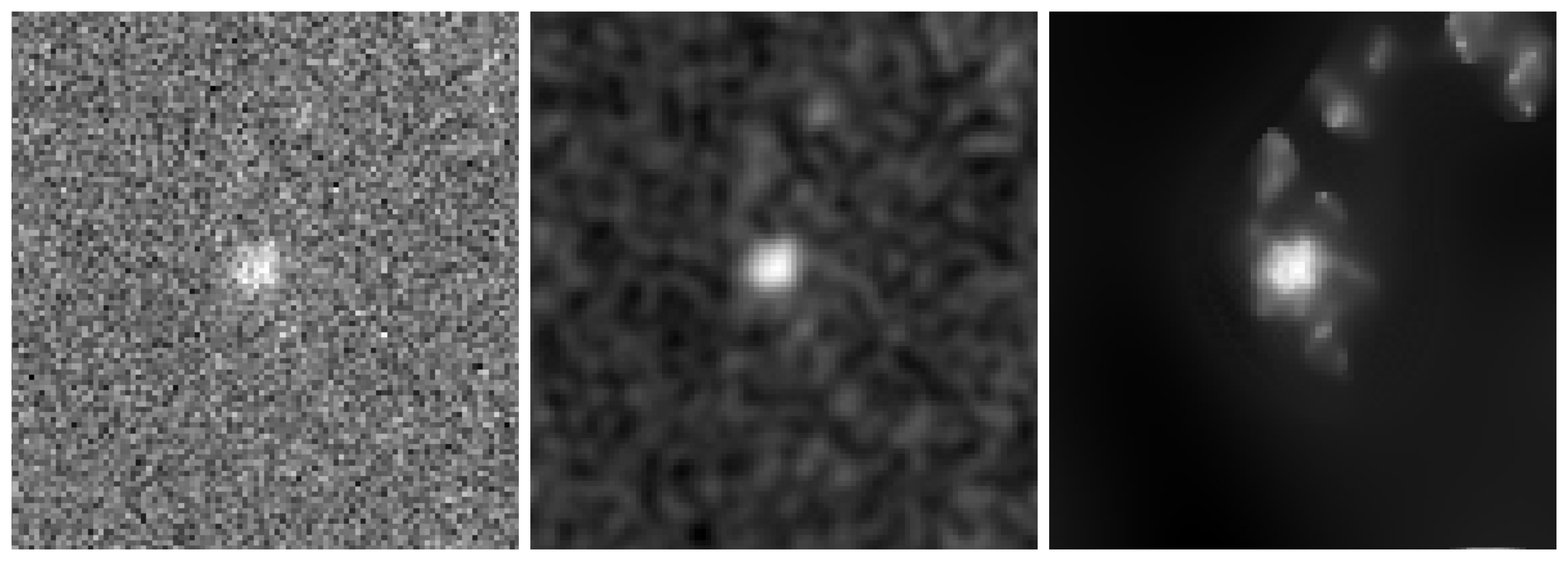}
\caption{From left to right: Raw CFHT $r$-band image; smoothed image; adaptively smoothed image of Mirabilis. DES detected the sources at the positions of adaptive smoothing revealed clumps, thus, the clumps are not artifacts of smoothing. By extension the smaller, fainter, secondary tail to the southern side is also real.
\label{fig:fig5}}
\end{figure*}
\subsection{Hot gas, XRB and ULX scenario \label{subsec:sub}}
In star forming galaxies the emission from hot interstellar-medium (ISM) is expected to contribute to the total X-ray emission. Although we do not have enough data to estimate these contributions, they are expected to account for a very small fraction of the observed X-ray emission. For example, a SFR of at least 30 \(\textup{M}_\odot\) yr$^{-1}$ is required to account for the faintest source in our work (X2) \citep{min10.1111/j.1365-2966.2012.21831.x}. \cite{Pardo_2016} studied 605 dwarf galaxies with stellar masses $<$3$\times$10$^9$ \(\textup{M}_\odot\), at $z<$1, and found that in all cases the SFR was $<$ 10 \(\textup{M}_\odot\) yr$^{-1}$, far from what would be required to account for X-ray detections from this paper. This implies that, unless all of our sources are extreme outliers, their X-ray luminosities are too high to be explained only by hot ISM gas and another source of X-rays is required. This is particularly clear in the case of Elstir's and Vinteuil's sources (X3 and X4) that have X-ray luminosities $>$ 10$^{42}$ erg s$^{-1}$. This implies that at least X3 and X4 are AGN, and that their infrared AGN-consistent colors are not due to star formation. Also, all four of our X-ray sources do not seem to be extended, as might be expected for diffuse gas or a spatially separated collection of X-ray binaries (XRBs). Furthermore, the faintest X-ray source (X2) is associated with a single clump (C1) in a long tidal tail. The other clumps have similar magnitudes, colors, and are probably made out of the same material and exposed to similar conditions and it is unlikely that a single clump is exhibiting extreme SFR, while the other clumps do not show similar behavior. \newline \indent
The X-ray sources associated with Mirabilis, X1 and X2, have luminosities just above 10$^{40}$ erg s$^{-1}$ and could be powered by a collection of stellar-mass accretors. Observations of Galactic X-ray binaries revealed that objects with X-ray luminosities of $>\times$10$^{38}$ erg s$^{-1}$ are quite rare \citep{grimmrefId0}. \cite{2019ApJS..243....3L} found that dwarf galaxies with stellar mass of 10$^9$ \(\textup{M}_\odot\), and SFR of 3 \(\textup{M}_\odot\) yr$^{-1}$ (which is ten times larger than what is typical for a dwarf starburst galaxy at $z<$0.3 \citep{Mezcua_2016}) are expected to have the integrated XRB luminosity of $\approx$10$^{40}$ erg s$^{-1}$. This is significantly lower than our least luminous X-ray source, X2. More up-to-date works studied low metallicity galaxies, a property typical for low-mass dwarfs, and found that contributions to the X-ray luminosity from the hot gas and high-mass X-ray binaries (HMXB) are significantly elevated compared to local relations \citep{2022ApJ...930..135L}. The derived relations for HMXB and hot gas contributions per unit SFR are log L$^{HMXB}_{0.5-8}$/SFR=40.19$\pm$0.06 and log L$^{gas}_{0.5-2}$/SFR=39.58$^{+0.17}_{-0.28}$, respectively. Using these elevated relations we conclude that the scenario in which the X-ray luminosity of the X1 source is entirely due to the population of HMXBs or X-ray emitting hot gas would require SFRs of 4 and 7 \(\textup{M}_\odot\) yr$^{-1}$, respectively. Although this SFR is not unheard of, it is still significantly larger than what is expected to be SFR of a dwarf starburst galaxy \citep{Mezcua_2016}, larger than almost all SFRs from \cite{2022ApJ...930..135L}, and larger than all SFRs from \cite{Pardo_2016}, \cite{Lemons_2015} and \cite{10.1093/mnras/staa040}. The X2 source requires lower SFRs for its X-ray emission to be explained with a population of HMXBs or hot gas, 2.5 and 4.5 \(\textup{M}_\odot\) yr$^{-1}$. Even though X2 requires lower SFRs than X1, these SFRs would still be one of highest ever reported in the dwarf galaxy related literature. Also, X2 is associated with a single clump (C1) in a long tidal tail, and it would be highly unusual for a single clump to host anomalously large population of luminous HMXBs or hot gas, while the other clumps do not show similar nature. We graphically represent this discussion in Figure \ref{fig:fig6}, where we show how our sources compare to elevated relations for X-ray luminosity produced by stellar processes, and to other dwarf galaxies from the literature.\newline \indent Next, we used hardness ratio (HR) as a proxy for the origin of the observed X-ray emission; obscured AGN usually have HR$>$0 (intrinsic N$_H >$10$^{22}$ cm$^{-2}$), unobscured AGN should have HR$\approx$--0.5, while HMXBs and hot X-ray emitting gas are expected to have HR$<$--0.8 \citep{Pardo_2016}. HR is defined as $\frac{H-S}{H+S}$,where H and S are number of counts in the hard (2-6 keV) and soft (0.3-2 keV) band, respectively. We find HRs for X1, X2, X3 and X4 to be --0.4$\pm$0.1, --0.3$\pm$0.2, --0.3$\pm$0.2 and 0.1$\pm$0.1, respectively. The X1, X2 and X3 sources are in line with being unobscured AGN, while the X4 source is in accordance with being an obscured AGN. We show these results in Figure \ref{fig:fig7}.\newline \indent Another possible explanation for our detections are ultraluminous X-ray sources (ULXs). ULXs are usually explained as stellar-mass black holes or neutron stars accreting at super-critical rates \citep{Begelman_2002} or objects with geometrically beamed radiation \citep{King_2001}. However, ULXs with X-ray luminosities $>$ 10$^{40}$ erg s$^{-1}$ are quite rare, and the Milky Way has none. \cite{2020MNRAS.498.4790K} found that 0.05 ULXs with L$_X>$ 10$^{40}$ erg s$^{-1}$ are expected to be found per \(\textup{M}_\odot\) yr$^{-1}$ of star-formation rate (SFR). A typical starburst dwarf galaxy at redshift $z<$0.3 is expected to have SFR=0.2 \(\textup{M}_\odot\) yr$^{-1}$ \citep{Mezcua_2016}, implying that the probability for a single starburst dwarf galaxy to host two ULXs with L$_X$ $>$ 10$^{40}$ erg s$^{-1}$ is P=0.00005, and several times smaller for L$_X$ $>$ 3$\times$10$^{40}$. Another work focused on ULXs in local dwarfs and found that galaxies with stellar mass M$_*$ $<$ 10$^{9}$ \(\textup{M}_\odot\) do not host any ULXs, which was explained by not having enough M$_*$ in their sample \citep{2008ApJ...684..282S}. Therefore, it is highly unlikely for a faint, small dwarf like Mirabilis to host two unusually bright ULXs. Obviously, the probability for the presence of a single ULX and a massive black hole is larger. However, even in this case this discovery can have significant importance for investigating how stellar mass black holes evolve with metallicity. The X-ray sources associated with Elstir and Vinteuil, X3 and X4 are too luminous to be ULXs, since their luminosities of $>$ 10$^{42}$ erg s$^{-1}$ are in accordance with luminosities of low-luminosity AGN.
\subsection{The Prospects of Adaptive Smoothing}
In recent years, an increasing number of works studied dwarf-dwarf galaxy mergers and effects these interactions have on galaxy evolution \citep{2015ApJ...805....2S,2020AJ....159..103K}. The key step, especially when studying isolated dwarfs, is detecting the tidal debris, which is an indicator of galaxy interactions. However, such tidal debris is expected to have very low surface brightness, thus challenging to detect. In this paper we showed that the application of adaptive smoothing can drastically improve searches for faint tidal features. Upon applying adaptive smoothing on our optical observations of average quality and depth, we were able to reveal tidal features with apparent magnitudes $>$25. Our findings open a new avenue in searching for tidal debris around dwarf galaxies and could possibly lead to discovering a rich population of interacting dwarf galaxies. \newline \indent
In order to illustrate the improvements we were able to obtain using adaptive smoothing, we show in Figure \ref{fig:fig5} the raw CFHT $r$-band image of Mirabilis, a smoothed image using a Gaussian function with radius of three, and an adaptively smoothed image using a Gaussian function, 15 counts under each kernel, with minimum and maximum kernel radii of 0.5 and 15, respectively, and logarithmic spacing between radii. The raw and Gaussian-smoothed images do not reveal any tidal debris, while the adaptively smoothed image clearly shows clumps of the main tail, but also uncovers the existence of a smaller, fainter secondary tail. The same effects were observed in the case of Elstir and Vinteuil, where the tidal bridge was only revealed upon carefully performing adaptive smoothing.
\section{Conclusions}\label{sec:fut}
In this paper we presented the current progress of our large-scale search for DAGN. We discovered two pairs of X-ray and infrared sources whose observed nature is in line with AGN. The host galaxies exhibit tidal features characteristic of merging/interacting galaxies and we conclude that they are most likely to be dwarfs. Therefore, the objects described in this paper represent strong candidates for the first DAGN in dwarf-dwarf galaxy mergers. We discuss possible alternate explanations for luminous X-ray sources, but we show that they are statistically highly likely to be AGN. Follow-up observations of these two systems will provide an opportunity to study various processes essential for understanding the earliest phases of the galaxy and black hole evolution, processes that are believed to be very common in the early Universe but remained hidden from us until now. We also demonstrate the importance of careful image processing in revealing low surface brightness tidal debris present in galaxy interactions. This paper introduces a novel approach of using archival X-ray and infrared observations, coupled with image processing to reveal DAGN in faint merging galaxies, and even if our targets turn out not to be dwarfs, the approach described in this paper can be used to discover rich populations of DAGN.
\begin{acknowledgments}
The authors gratefully acknowledge the use of \textsc{CIAO} software, as well as the Ned Wright Cosmology Calculator. MM and JAI are supported by Chandra Grant AR2-23004X. Authors thank Preethi Nair for useful discussions.
\end{acknowledgments}
\bibliography{sample631}{}
\bibliographystyle{aasjournal}
\end{document}